\pdfoutput=1
\documentclass[11pt]{article}
\usepackage{amsfonts,amsmath,amsthm,array,amssymb}
\usepackage{cite, graphicx, geometry, subfigure}
\usepackage{url, units}
 \usepackage{hyperref}
\usepackage{epstopdf}
\usepackage{centernot}
\usepackage{subfigure}
\usepackage{multicol}

\makeatletter
\newcommand\appendix@section[1]{%
  \refstepcounter{section}%
  \orig@section*{Appendix \@Alph\c@section: #1}%
  \addcontentsline{toc}{section}{Appendix \@Alph\c@section: #1}%
}
\let\orig@section\section
\g@addto@macro\appendix{\let\section\appendix@section}
\makeatother
\begin{document}
\pdfoutput=1
\title{Short and Long Range Population Dynamics of the Monarch Butterfly (\emph{Danaus plexippus})}
\author{Komi Messan$^1$, Kyle Smith$^2$, Shawn Tsosie,$^3$ Shuchen Zhu$^4$, Sergei Suslov$^5$\\
\footnotesize{$^1$ komimessan@gmail.com, $^2$ ksmi02@nmt.edu, $^3$ tsosie@mit.edu, $^4$ szhu11@asu.edu, $^5$ sks@asu.edu} }
\date{\today}
\maketitle

\begin{abstract}
The monarch butterfly annually migrates from central Mexico to southern Canada.
During recent decades, its population has been reduced due to human interaction with their habitat.
We examine the effect of herbicide usage on the monarch butterfly's population by creating a system of linear and non-linear ordinary differential equations that describe the interaction between the monarch's population and its environment at various stages of migration: spring migration, summer loitering, and fall migration.
The model has various stages that are used to describe the dynamics of the monarch butterfly population over multiple generations.
In Stage 1, we propose a system of coupled ordinary differential equations that model the populations of the monarch butterflies and larvae during spring migration.
In Stage 2, we propose a predator-prey model with age structure to model the population dynamics at the summer breeding site.
In Stages 3 and 4, we propose exponential decay functions to model the monarch butterfly's fall migration to central Mexico and their time at the overwintering site.
The model is used to analyze the long-term behavior of the monarch butterflies through numerical analysis, given data available in the research literature.
\end{abstract}

\section{Introduction}
The migration of the monarch butterfly is a marvel of nature.
It is a journey through time and space that spans a distance of 4500 km and at least four generations of monarch butterfly, the exact number of generations depends on local climatological conditions \cite{thermo1988}.
The persistence of this yearly cycle is heavily dependent on two plants: the milkweed plants, of the family (\emph{Asclepiadaceae}) found all over North America, and the Oyamel fir tree found at its overwintering habitat \cite{book4}.
The milkweed is the only food source for the larvae and the Oyamel fir trees help keep the monarch in a cool state during its winter hibernation \cite{thermo1988}.
Due to deforestation in the mountains of central Mexico and the increased usage of herbicide in the United States and Canada, the plants the monarch butterfly depends on have been reduced \cite{Brower2011}.

We are primarily concerned with the effect herbicide usage has on the long-term population dynamics of the monarch butterfly.
For this purpose, we develop a multi-stage model that describes the monarch butterflies migration and use this information to create a discrete time model of the behavior of the population year after year.
Our model is based on values obtained from previous research on the monarch butterfly migration and the milkweed plant.
We then use our model to estimate the impact of herbicide usage on the population of monarch butterflies.

\section{Biological Background}

\subsection{Monarch Butterfly Life-Cycle}

\begin{figure}
\centering
\includegraphics[width=7cm]{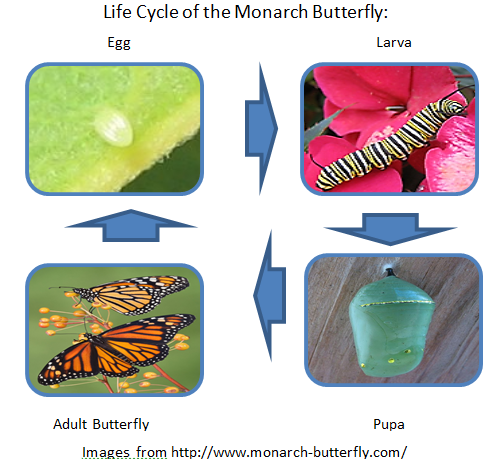}
\caption{The life-cycle of the monarch butterfly has four stages: egg, larva, pupa and adult}
\end{figure}

The monarch butterfly (\emph{Danaus plexippus}) is rare among migratory animals and unique among insects.
In the family of insects, the desert locust is the only other species that migrates a comparable distance \cite{desertlocustuf}.
Desert locusts have a dynamic migratory cycle, a cycle dependent on ``directed movement controlled by tides or wind, with navigation abilities not essential'' \cite{desertlocustuf}, unlike the monarch butterflies, which have a seasonal migration.
It is rare among migratory animals, because the generation that leaves the overwintering site in central Mexico, in the spring, is not the generation that returns to the overwintering site the following fall \cite{book4}.
There are multiple subspecies of monarchs, migratory and nonmigratory \cite{geneticclass2005}.
A subspecies known as \emph{Danaus plexippus plexippus} are the migratory monarchs, which will be the primary focus of this paper.
There are several migratory populations of monarch butterflies as well; migrant monarch butterflies that live east and west of the Rocky Mountain range.
We focus primarily on the populations east of the Rocky Mountain range, because they have the largest population and the longest migration route \cite{ref9}.

There are several stages of the monarch butterflies flight: the spring migration from central Mexico to southern Canada, the summer loitering in southern Canada, and the fall migration from southern Canada back to central Mexico \cite{Reppert2010}.
We will follow the convention of Lincoln P. Brower and designate the monarch populations that are traveling, either from south to north or from north to south, as migrants \cite{Brower1996}. 

The monarch butterfly migration begins in the Oyamel fir trees on the mountain of Sierra Palon in central Mexico \cite{Brower1996}, where they spend the winter in a state of torpor and reproductive diapause, a state of non-reproduction \cite{ref9}.
Research has shown that shorter day length, lower temperature, and larvae feeding on older milkweed increases the likelihood that a monarch will enter reproductive diapause, the state necessary for fall migration \cite{diapause2002}.
Though the mechanisms are known, the exact cause of the monarch butterfly's transition to its autumnal migratory state is unknown and is a current subject of research \cite{Reppert2010} .  

When the unknown mechanism is triggered, the monarch butterflies begin their migration, leave reproductive diapause, and become reproductively active \cite{diapause2002}.
While reproductively active, the female monarch can lay up to 700 eggs during her lifespan of seven to nine months \cite{book4}.
Female monarchs search for young milkweed leaves to lay their eggs, laying one egg per milkweed leaf, before flying off to find another milkweed plant to lay more eggs \cite{book4}.
After laying an egg, the female monarch resumes her flight north, continually laying eggs until she dies.  

Meanwhile, the offspring hatch from their eggs after three to four days  \cite{book4}.
The larvae begin life by consuming portions of their egg before moving on to eat the milkweed plant \cite{book4}.
The larvae have five stages of growth, called instars.
The first four instars end after each larval molt and the final instar ends when the larvae become pupae.
The complete larval stage lasts approximately two weeks, where the larvae spend the entire stage on one milkweed, during which the larva grows to about 2500 times its original size \cite{book4}.  

The monarch larvae search for a dark place to begin their pupal stage.
This stage lasts approximately ten days, during which the entire structure of the larvae breaks down to be reconstituted into their adult form \cite{book4}.
Urquhart noted that temperature can either retard or accelerate the growth rate at every stage of development in the monarch butterfly life cycle.
This means that the ten days given for the monarch pupal stage, like the other values, are averages.

At the end of the pupal stage, the larvae become adult monarchs ready to resume the migration begun by its parents.
Unlike its parents, its life is reduced by a significant amount, living only two to six weeks, whereas its overwintering parents lived up to nine months \cite{book4}.
We simulate this part of the life-cycle of the monarch butterfly in \emph{Stage 1} of our model.
The migration continues in this fashion, parents beget larvae, the parents die, the larvae grow up and fly further north.

Lincoln P. Brower determined the geographical extent of each generation of monarch butterflies through chromatography analysis of the cardenolides, the toxic chemical found in milkweed plants, inside each monarch butterfly \cite{Brower1996}.
The cardenolides in different species of milkweed plants have specific chemical profiles and each of these milkweed species is located within different geographic ranges.
Brower used these two facts to determine that ``the first spring generation is produced largely in Texas and Louisiana'' and ``continue the migration northwards to the Great Lakes region and Southern Canada'' \cite{Brower2006}.

We model the next phase of the migration, the time the monarch stays in Southern Canada, in \emph{Stage 2} of our model.
This spring migration usually begins in the middle of March and ends in early June \cite{book4}.

The monarch butterflies continue their life-cycle in Southern Canada and the Northern United States.
This stage usually lasts from mid-June to mid-August.
At the end of this stage, the monarch receives environmental cues that cause it to enter reproductive diapause \cite{book4}.
The monarch butterfly becomes a migrant and begins its return south toward the overwintering sites of the previous fall.
We simulate this behavior in \emph{Stage 3} of our model.
When the monarch enters reproductive diapause, it increases its lipid stores by constantly feeding on nectar \cite{Brower2006}.
The monarch butterfly needs this lipid reserve to survive the winter, during which it will feed at a much reduced rate \cite{Brower2006}.
Once migrants arrive to central Mexico they enter a state of hibernation which we simulate with \emph{Stage 4} of our model.
Unlike the spring migration which is composed of multiple generations of migrants, the fall migration consists of only one generation of migrants \cite{ref9}.

\subsection{Common Milkweed (\emph{Asclepiadaceae syriaca}) Life-Cycle}

An understanding of the monarch butterfly life-cycle would be incomplete without some discussion of the milkweed family (\emph{Asclepiadaceae}) of plants.
Milkweed plants produce cardenolides (cardiac glycosides) and latex \cite{agrawal} as their primary defensive measures.
We primarily focus on the life-cycle and development of the common milkweed plant and the effect of herbicide on its development.
The common milkweed is a perennial plant that reproduces primarily from shoots off the main plant colony, but also reproduces from seeds \cite{Holdrege}. 
It typically grows between 1 and 1.5 meters tall \cite{Holdrege}.
The young leaves are the preferred site for the monarch female to lay her eggs \cite{book4}.
 
The cardiac glycosides found in the common milkweed are toxic to many animals, because it ``can disrupt the ionic balance of a number of different cell types in animals, including heart muscle, vascular smooth muscle, neurons, and kidney tubules'' \cite{Holdrege}.
An indication of the level of destruction of the milkweed by herbicide can be found in Iowa.
In 1999, the common milkweed was present in approximately 50\% of Iowa corn and soybean fields.  In 2009, the percentage of common milkweed was present in only 8\% of the fields \cite{Hartzler20101542}. 
Since much of the Midwest is farmland, this is an indication that a significant portion of the monarch butterfly habitat is at risk.

\subsection{Previous Mathematical Work}

Since $1960$, Lincoln Brower, Fred Urquhart, other zoologists, and other biologists have tried to understand the life-cycle and migration of the monarch butterfly.
Despite all the work conducted on monarch butterflies from a zoological and biological perspective, there has been a dearth of mathematical work.
To our knowledge the only other work was a discrete model by Yakubu et al. \cite{Yakubu2004}.
The authors modeled the migration and life cycle of the monarch butterfly with a set of difference equations \cite{Yakubu2004}.
They assume that the spring migration, from the overwintering site in Mexico to Southern Canada, consists of three generations and the fall migration, which travels from southern Canada back to Mexico and consists of one generation.
This comprises a total of 4 generations throughout the entire cycle.
The goal of their project was to investigate the persistence of the monarch butterfly population, with a spatially discrete advection model with emphasis on compensatory (contest competition) and overcompensatory (scramble competition) dynamics \cite{Yakubu2004}.

In their work, the authors assumed non-stochastic extinction of the population and discrete reproduction during the spring migration.
A threshold parameter, or basic reproductive number, for the persistence or extinction of the monarch butterfly population was found and they analyzed it in different situations.
Based on their findings, extinction or persistence of the population in generation 4 depends on the non-migratory population size in generation 3.
Different behavior is observed with different parameters and non-migratory population size.

In our model, we assume that during their migration, the monarch butterflies reproduce continuously.
We also model their movement by advection.
Finally, we include the major larval food source, the common milkweed. We numerically investigate the long-range behavior (over 30 years), of our model. 

\section{A Model Under Consideration}

\subsection{Description of the Model}

\begin{figure}[htp]
\centering
\subfigure[2011 monarch butterfly spring migration]
{%
\includegraphics[width=7cm]{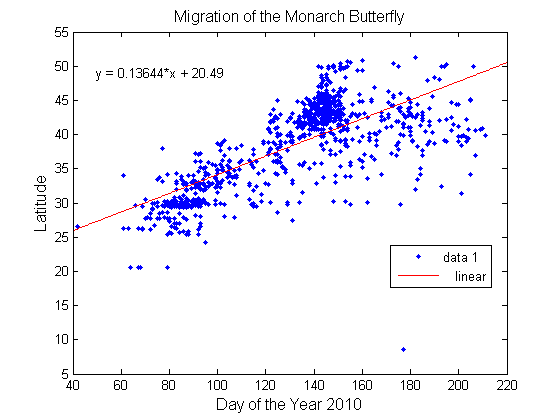}
\label{fig:line}
}%
\subfigure[Temperature Data for First Monarch Sightings by Day (2010)]
{
\includegraphics[width=7cm]{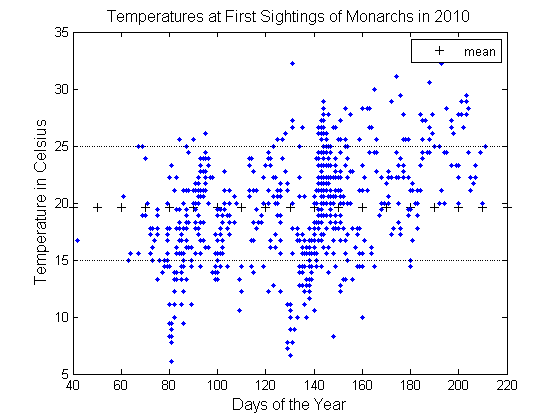}
\label{fig:monarchtemp}
}
\caption{}
\end{figure}

The difficulty of modeling the monarch butterfly arises from its unique migratory nature.
We analyzed data from the website Journey North \cite{jnorth} and found that monarch butterfly stays within a temperature range of approximately 15.5 $^{\circ}$C to 24 $^{\circ}$C, see Figure \ref{fig:monarchtemp}.
We also analyze the first monarch sightings of the year, available on the website Journey North \cite{jnorth}, and found that monarch butterflies travel at an approximately linear rate, see Figure \ref{fig:line}.
With this, we assume that the amount of milkweed available to the monarch butterfly is constant, because the monarch butterfly constantly moves north into areas of new milkweed.
Thought of in another way, we assume that the monarch butterflies occupy an ``expanding  box" traveling at a constant speed north.

We divide the model of the migration cycle into four stages.
The \emph{first stage} focuses on the monarch migrants' flight from central Mexico to southern Canada.
This stage incorporates multiple generations, since the monarch butterfly continually reproduce along the way.
We consider the initial generation as generation zero, those monarch butterflies that survive the hibernation phase in central Mexico.
We consider generation $i$ as the monarch butterfly that arrive in southern Canada (according to existing data over the years, $3 \leq i \leq 7$).

The \emph{second stage} describes the butterfly in southern Canada and takes into consideration the effect of herbicide on the common milkweed.
We assume an age-structure on the monarch butterfly, larva and adults.
We also assume a predator-prey model, with the larvae as the predator and the milkweed as the prey.

The \emph{third stage} of the model simulates the monarch migrants' return to central Mexico from southern Canada.
We model this with an exponential decay function, because the monarch butterfly is non-reproductive during the fall migration, yet continually die due to various environmental factors (weather, natural catastrophes, etc. $\ldots$ \cite{book2}).

The \emph{fourth stage} models the hibernation phase and we also simulate this stage with an exponential decay function, because the monarch butterflies die during their dormant phase.

\subsection{Stage 1}
Our model for the \emph{first stage} is presented in the following form and the parameters are shown in Table \ref{table $1$:hParameters}. 

\begin{figure}
\centering
\includegraphics[width=13cm]{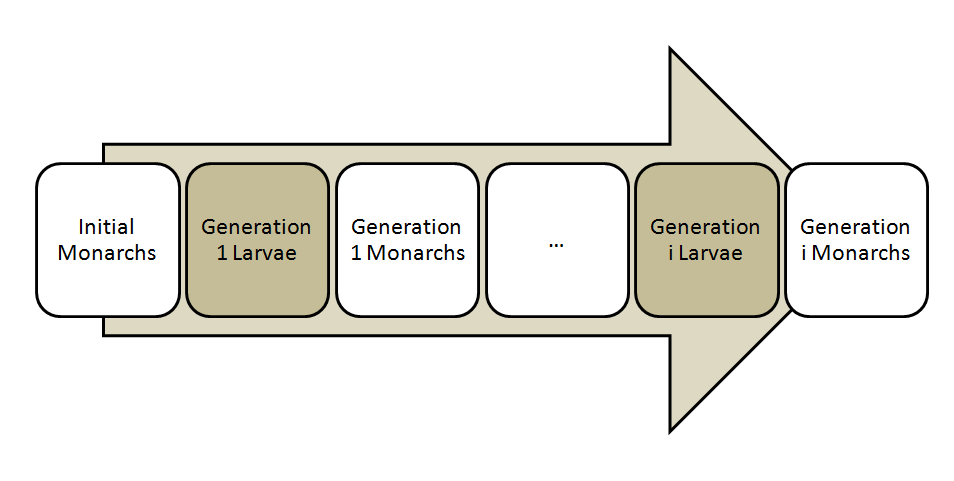}
\caption{Flow of monarch butterfly reproduction from first to last generation in \emph{Stage 1}.}
\label{stageoneimg}
\end{figure}

We propose the following system of equations to model the population dynamics of the monarch butterfly during the spring migration:
\begin{align}
\frac{d M_{0}}{dt} 	& = -\mu_{0}M_{0}, \label{stageoneeqone}\\
\frac{d L_{1}}{dt} 	& = \alpha_{1} M_{0} A_{0} - \left(\gamma + \mu_{1}\right) L_{1}, \label{stageoneeqtwo}\\
\frac{d M_{1}}{dt}	& = \gamma L_{1} - \mu_{2} M_{1}, \label{stageoneeqthree}\\
\frac{d L_{2}}{dt} 	& = \alpha_{1} M_{1} A_{0} - \left(\gamma + \mu_{1}\right) L_{2}, \label{stageoneeqfour}\\
\frac{d M_{2}}{dt} 	& = \gamma L_{2} - \mu_{2} M_{2}, \label{stageoneeqfive}\\
\vdots \notag \\
\frac{d L_{i}}{dt} 		& = \alpha_{1} M_{i-1} A_{0} - \left(\gamma + \mu_{1}\right) L_{i}, \label{stageoneeqsix}\\
\frac{d M_{i}}{dt} 	& = \gamma L_{i} - \mu_{2} M_{i}, \label{stageoneeqseven}
\end{align}
with initial conditions
\begin{equation*}
M_{0}(t_{0}) \neq 0 \qquad L_{j}(t_{0}) = M_{j}(t_{0}) = 0 \, \textrm{for} \, 1 \leq j \leq i,
\label{initone}
\end{equation*}
and with the parameters given in Table \ref{table $1$:hParameters}.
\begin{table}[htp]
\caption{Parameters for Stage 1 (for more details see Appendix \ref{appstageone})}
\centering 
\begin{tabular}{l l l l} 
Parameter & Biological Meaning & Default Value & Unit\\ 
\hline 
$\mu_{0}$   	&  Death rate of overwintering monarchs 		& 0.1198 -- 0.1997 	& $\mathrm{1/day}$\\ 
$\alpha_{1}$	& Growth rate of larvae	 				& 2.232 			& $\mathrm{1/day}$ \\ 
$A_{0}$ 		& Percentage milkweeds not killed by herbicide 	& 1 				&\\
$\gamma$		& Maturation rate of larvae 				& 0.03571 			& $\mathrm{1/day}$  \\ 
$\mu_{1}$ 		& Death rate of larvae 					& 0.0902 -- 0.1397 	& $\mathrm{1/day}$  \\ 
$\mu_{2}$ 		& Death rate of adult monarchs			& 0.07143 			& $\mathrm{1/day}$  	
\end{tabular}
\label{table $1$:hParameters}
\end{table}
In Equation \ref{stageoneeqone}, $M_{0}$ represents the monarch butterfly population in central Mexico.
In Equations \ref{stageoneeqtwo} and \ref{stageoneeqthree}, $L_1$ and $M_1$ represent the population of the first generation larvae and monarch butterflies who migrate to Canada in the spring.
In Equations \ref{stageoneeqfour} and \ref{stageoneeqfive}, $L_2$ and $M_2$ represent the populations of larvae and monarch butterflies, in the second generation who migrate to Canada, respectively.
Similarly in Equations \ref{stageoneeqsix} and \ref{stageoneeqseven}, from the model $L_i$ and $M_i$ are the population of larvae and monarch butterflies in the $i^{th}$ generation. 

In \emph{Stage one} Equation \ref{stageoneeqone}, the term $-\mu_{0} M_{0}$ describes the rate of change in the population of fall migrants, the monarch butterflies that overwinter in central Mexico.
The term $\mu_0$ is the death rate of the initial generation of monarch butterflies.
These monarch butterflies were in reproductive diapause and they will not produce monarch butterflies who are in reproductive diapause, so the population decreases at a proportional rate.

In Equation \ref{stageoneeqtwo}, the parameter $\alpha_1$ is the per capita growth rate of the larvae,given in Table \ref{table $1$:hParameters}.
The rate $\alpha_{1}$ is multiplied by the initial population leaving Mexico and the amount of milkweed, which is a ratio, along the way since the larvae depend on the milkweed.
Thus, the amount of larvae, $L_{1}$, is proportional to the milkweed, $A_{0}$.
The second term $\gamma$ in Equation \ref{stageoneeqtwo} is the maturation rate of the monarch butterfly, from larva to adult, so we have the term $-\gamma L_{1}$.
We also consider those larvae that die before they become adults.
They leave the population, so we have the term $-\mu_{1} L_{1}$.

Next, we look at Equation \ref{stageoneeqthree} in stage one.
$\gamma$ is the maturation rate of larvae. 
The monarch butterflies also die at a rate proportional to their population, thus we have $-\mu_{1}$ multiplied by $M_{1}$, the number of monarch butterflies in generation one.
This pattern continues until the $i^{th}$ generation (the total number of reproduction generations in stage one), when the monarch butterfly population reaches southern Canada.

\subsection{Stage 2}
\begin{figure}
\centering
\includegraphics[width=13cm]{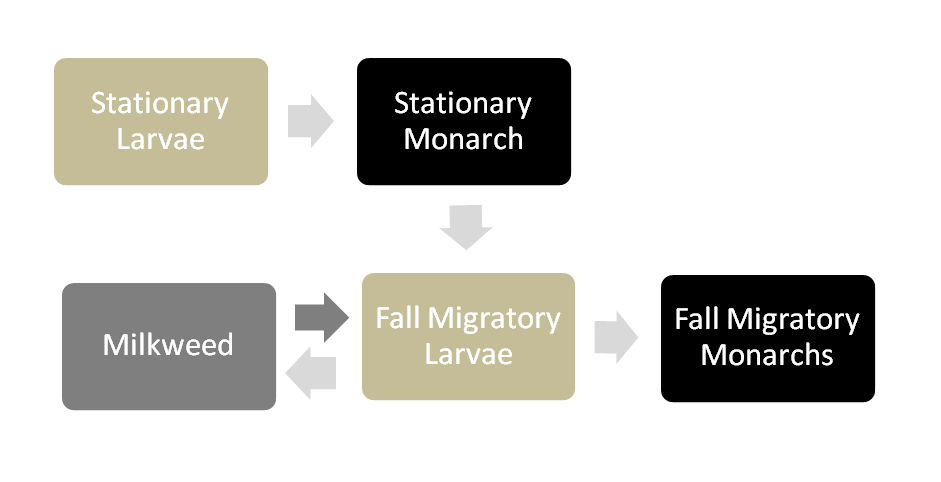}
\caption{Interaction between larvae, adult, and milkweed.}
\label{stagetwoimg}
\end{figure}

We model stage two with the following system of equations:
\begin{align}
\frac{d L_{in}}{dt} 	& = -(\gamma + \mu_{1}) L_{in}, \label{stagetwoeqone}\\
\frac{d M_{in}}{dt} 	& = -\mu_{2}M_{in} + \gamma L_{in}, \label{stagetwoeqtwo}\\
\frac{d L_{s}}{dt} 	& =  \alpha_{2} A M_{in} - \left(\gamma + \mu_{1}\right)L_{s}, \label{stagetwoeqthree} \\
\frac{d M_{s}}{dt} 	& = \gamma L_{s} - \mu_{3} M_{s}, \label{stagetwoeqfour} \\
\frac{d A}{dt} 		& = aA \left(1 - \frac{A}{K}\right) - A (\sigma + \beta L_{s}). \label{stagetwoeqfive}
\end{align}
The previous equations have the initial conditions:
\begin{equation*}
\label{inittwo}
A(t_{1}) \neq 0, \qquad L_{s}(t_{1}) = M_{s}(t_{1}) = 0, \qquad M_{in}(t_{1}) = M_{i}(t_{1}), \qquad L_{in}(t_{1}) = L_{i}(t_{1}).
\end{equation*}
and the parameters listed in Table \ref{tab:hresult}.
\begin{table}[h]
\caption{Parameters for Stage 2 (for more details see Appendix \ref{appstagetwo})} 
\centering 
\begin{tabular}{l l l l} 
Para.& Biological Meaning & Parameter Value & Unit\\    
\hline             
$\gamma$		& Maturation rate of Larvae 				& 0.03571 			& $\mathrm{1/day}$  \\ 
$\mu_{1}$ 		& Death rate of Larvae 					& 0.0902 -- 0.1397 	& $\mathrm{1/day}$  \\ 
$\mu_{2}$ 		& Death rate of later monarchs				& 0.07143 			& $\mathrm{1/day}$  	\\
$\alpha_{2}$	& Growth rate of larvae with Monarchs		& 2.6 				& $\mathrm{m^2}/(\mathrm{kg*day})$ \\ 
$\mu_{3}$ 		& Death rate of non-reproductive monarchs		& 0.005 			& $\mathrm{1/day}$  	\\
$a$   			& Growth rate of milkweed 				& 0.007 			& $\mathrm{1/day}$\\ 
$K$			& Carrying capacity of milkweeds			& 1.79188 			& $\mathrm{kg}/\mathrm{m^2}$ \\ 
$\sigma$ 		& Percentage of milkweed destroyed by herbicide 	& 1 				& $\mathrm{1/day}$ \\
$\beta$		& Consumption of milkweed by larvae		& $5 \cdot 10^{-9}$ 	& $1/ \mathrm{(larvae* day)}$ 
\end{tabular}
\label{tab:hresult}
\end{table}
In \emph{Stage 2}, we consider the interaction between the larvae, the milkweed, and the adult monarch butterflies and also consider the effect of herbicide on the milkweed.
We obtain the visual illustration of the model in Figure \ref{stagetwoimg}.

The terms $L_{in}$ and $M_{in}$ are the population of larvae and monarch of the last generation of stage one, respectively.
The terms $L_{s}$ and $M_{s}$ are the population of larvae and monarch adults that are in reproductive diapause.
A small portion of the larvae mature and become adult butterflies and this is reflected by the term $-\gamma L_{in}$ in Equation \ref{stagetwoeqone}, where $\gamma$ is the maturation rate, see Table \ref{tab:hresult}.
The parameter $\mu_{2}$ is the mortality rate of the adult monarch butterfly.
Equation \ref{stagetwoeqtwo} represents the population increase of reproductively active monarch butterflies.

The larvae of $M_{in}$ are denoted by $L_{s}$ and they either die or they mature, becoming $M_{s}$.
We model this with Equations \ref{stagetwoeqthree} and \ref{stagetwoeqfour} of \emph{Stage two}.
Equation \ref{stagetwoeqfive}, describes the interaction between the milkweed, the larvae, and the herbicide.
This system is a modified \emph{predator-prey} system where the larvae are the predator and the milkweed is the prey.
Adding the adult butterfly to the system changes the dynamics of the system, since the butterfly has a positive effect on the plant through pollination \cite{book4}.
Since there are multiple pollinators of the milkweed, we consider the effect of the adult butterfly negligible.

\subsection{Stage 3}

In the \emph{third stage}, the butterflies are migrating back to central Mexico and are in reproductive diapause, therefore their populations doesn't increase.
Also, a significant proportion of them do not reach central Mexico due to the following factors: natural catastrophes, weather, and other environmental factors \cite{book2}.
Thus, we have an exponential decay function of the following form, where $\mu_{fin}$ represent the death rate at this stage.
The number of butterflies in reproductive diapause that leave southern Canada, $M_{s}(t_{2})$, are the initial condition of the following equation:
\begin{align}
\frac{d M_{fin}}{dt} &= -\mu_{fin} M_{fin}.
\end{align}

\subsection{Stage 4}

The last stage models the butterfly in its dormant phase.
This is qualitatively similar to \emph{Stage 3}, because they are still in reproductive diapause, and a large portion of them will die.
We model this with an exponential decay function using a different rate than \emph{Stage 3}.
The population of butterfly that arrive in central Mexico, from \emph{stage three}, represent the initial condition, $M_{w}(t_{3})$, of the following equation:
\begin{align}
\frac{d M_w}{dt} = -\mu_{w} M_{w}.
\end{align}

\begin{table}[htp]
\caption{Parameters for Stage 3 and 4 (for more details see appendix \ref{appstage3and4})} 
\label{table:3and4parameters}
\centering 
\begin{tabular}{llll}
Parameter & Biological Meaning & Default Value &Unit \\    
\hline            
$\mu_{fin}$ 		& Death rate of migrant monarchs			& 0.0056 	& $\mathrm{1/day}$  \\ 
$\mu_{w}$ 			& Death rate of overwintering monarchs		& 0.0042 & $\mathrm{1/day}$  	\\
\end{tabular}
\label{tab:stage3and4}
\end{table}

\section{Numerical Results}

\subsection{Preliminary Results}

From the data we obtain from the Journey North \cite{jnorth}, we see a correlation in the distribution of milkweed and monarch butterfly as they travel from the overwintering site in Mexico to southern Canada.

\begin{figure}%
\centering
\subfigure[2011 monarch butterfly spring migration]
{%
\includegraphics[width=7cm]{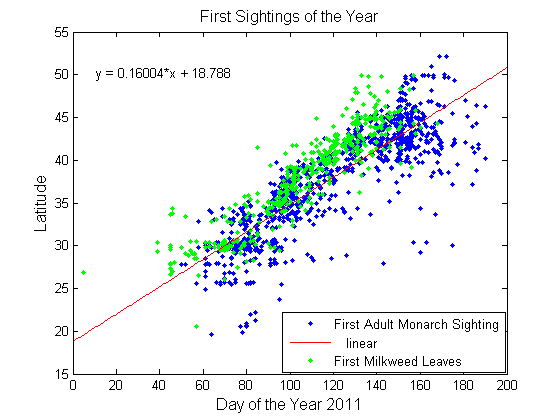}
\label{fig:monmig}
}%
\subfigure[2010 monarch butterfly fall and spring migration \cite{jnorth}]
{
\includegraphics[width=7cm]{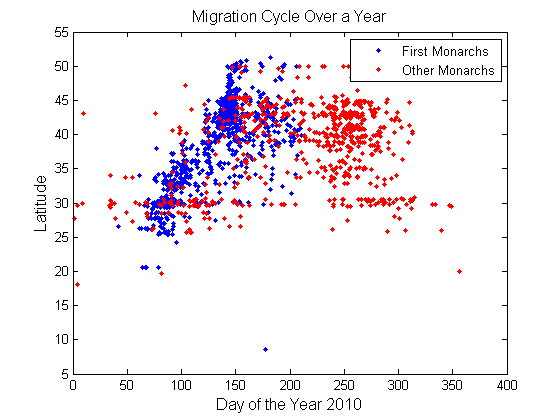}
\label{fig:monarchsightings}%
}%
\caption{First Monarch Sightings (2010) \cite{jnorth}}
\end{figure}

Figure \ref{fig:monmig} shows the distribution of both new leaves and adult butterflies throughout the United States, with respect to different latitudes (from the southern to northern sections of United States).
Again with data from Journey North \cite{jnorth}, we obtain a similar illustration for other monarch butterfly sightings for both the fall and spring migrations, the complete annual migration, as shown in Figure \ref{fig:monarchsightings}.

\subsection{One Year}

\begin{figure}%
\centering
\subfigure[Annual monarch butterfly population behavior in the United States with $A_0 = 0.6$ and $\sigma=1$ per day.]{%
\includegraphics[width=7cm]{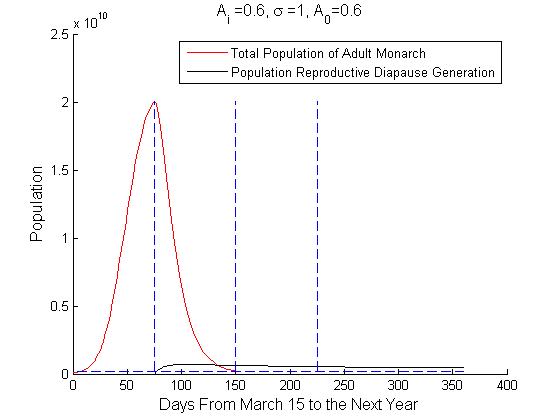}
\label{fig:zeropointsix}%
}
\hspace{8pt}%
\subfigure[Annual monarch butterfly population behavior in the United States with $A_0 = 0.3$ and $\sigma=1$ per day.]{%
\includegraphics[width=7cm]{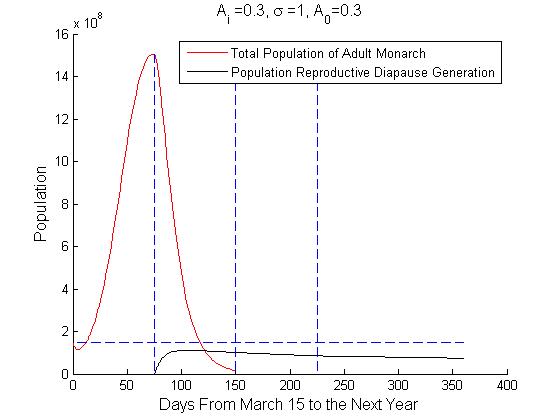}
\label{fig:zeropointthree}%
}\\
\subfigure[Simulation of the monarch butterfly population in the United States over a $30$ year period with $A_0 = 0.6$ and $\sigma=1$ per day.]{%
\includegraphics[width=7cm]{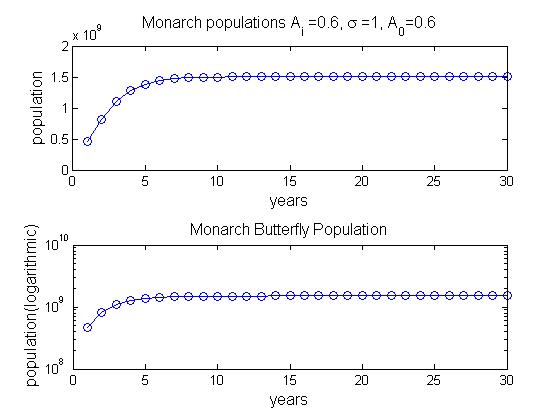}
\label{fig:zeropoinsixandone}%
}%
\hspace{8pt}%
\subfigure[simulation of the monarch butterfly population in the United States over a $30$ year period with $A_0=0.3$ and $\sigma=1$ per day.]{%
\includegraphics[width=7cm]{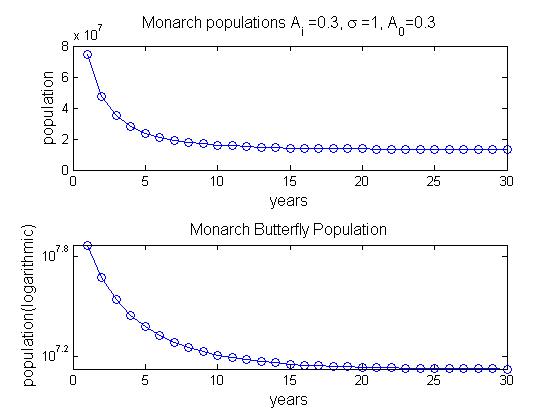}
\label{fig:zeropointthreeandone}%
}%
\caption{Variation of the amount of milkweed along the migratory route.}%
\label{fig:diffina}%
\end{figure}

In this section, we look at the behavior of the monarch butterfly population over one year, under various conditions.
We start with an initial monarch population of $150,000,000$, based on the approximation found on the Journey North website \cite{jnorth}, an initial milkweed percentage of $0.6$ and $0.4$, and we use the various parameters in Tables \ref{table $1$:hParameters} through \ref{table:3and4parameters}.
We generate the annual population behavior shown in Figures \ref{fig:zeropointsix} and \ref{fig:zeropointthree}.

\begin{figure}%
\centering
\subfigure[Annual monarch butterfly population behavior in the United States with $A_0=0.4$ and $\sigma=10$ per day.]{%
\includegraphics[width=7cm]{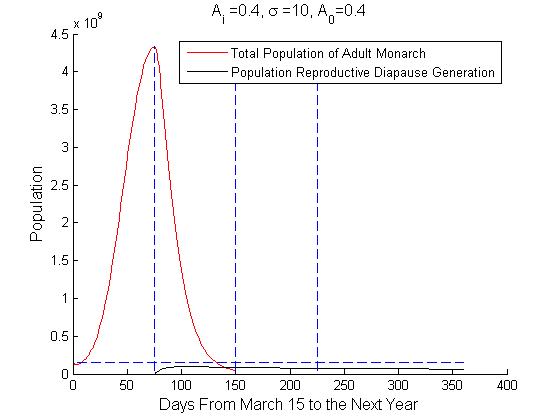}
\label{pointfourandten}%
}
\hspace{8pt}%
\subfigure[Annual monarch butterfly population behavior in the United States with $A_0=0.4$ and $\sigma=1$ per day.]{%
\includegraphics[width=7cm]{DifInA3.jpg}
\label{pointfourand1}%
}\\
\subfigure[Simulation of the monarch butterfly population in the United States over a $30$ year period with $A_0=0.4$ and $\sigma=10$ per day.]{%
\includegraphics[width=7cm]{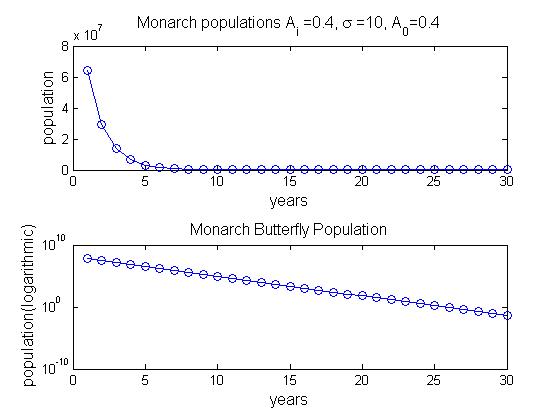}
\label{pointfourandten2}%
}%
\hspace{8pt}%
\subfigure[Simulation of the monarch butterfly population in the United States over a $30$ year period with $A_0=0.4$ and $\sigma=1$ per day.]{%
\includegraphics[width=7cm]{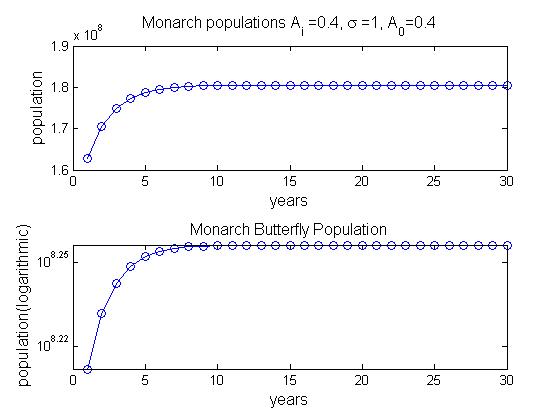}
\label{fig:pointfourandone}%
}%
\caption{Variation of $\sigma$ for various values of $A_{i}$.}%
\label{fig:diffinsig}%
\end{figure}

Next, we run the simulation with different values of $\sigma$, the herbicidal rate.
As we see in Figures \ref{fig:zeropointsix} through \ref{fig:pointfourandone}, we have different behavior when we vary the value of $\sigma$.
First, we consider $A_0$ (initial milkweed percentage in kg per square meter) equal to $0.4$ and $\sigma$ equal to 10 and we obtain the graphs in Figure \ref{fig:diffina}.

\subfiglabelskip=0pt

When simulate our model over a year we observe unique dynamic behaviors.
The generations of the monarch butterflies overlap grow over time.  When they arrive at Canada we observe a different behavior.
There is a decline in the monarch and larvae population.
When this simulation is repeated over many years the butterfly population at the end of each cycle exhibits discrete logistic-type behavior.

If the constants $A_{0}$ and $A(t_{1})=A_{i}$ are sufficiently small, we observe a change in behavior.
The graphs of the population, with respect to time, are inverted.  This suggests a critical point in the graphs.
When the simulations are run close to the turning point we note that the range of values becomes decreases.
We found the critical point numerically and it's value is approximately $A_{0} = A_{i} = 4.019$.
When looking at this over one year, it we see that a higher value of $A$ will show logistic growth and a lower value of $A$ will show logistic decay, depending on how extensive the herbicide is used.  

\subfiglabelskip=0pt

When we vary $\sigma$ in the equations we observe that if $\sigma$ is small, then the population is more localized in one area.  In fact, increasing $\sigma$ by a factor of 10 eventually leads to extinction, while smaller values of $\sigma$ shows stabilization.

\begin{figure}%
\centering
\subfigure[Annual monarch butterfly population behavior in the United States with $A_0 = 0$ and $\sigma = 100000$ per day.]{%
\includegraphics[width=7cm]{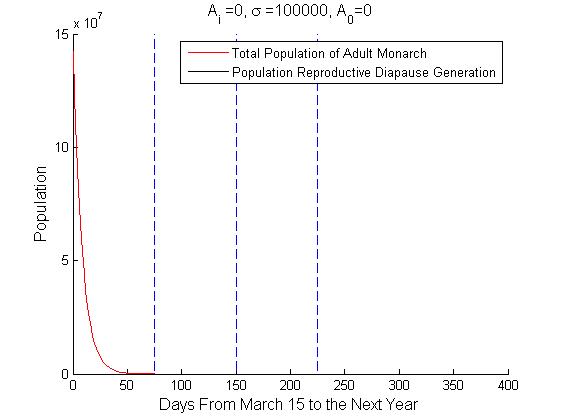}
\label{deadbutterflies}
}
\hspace{8pt}%
\subfigure[Prediction of the monarch butterfly population in the United States over a $30$ year period with $A_0 = 0$ and $\sigma=100000$]
{
\includegraphics[width=7cm]{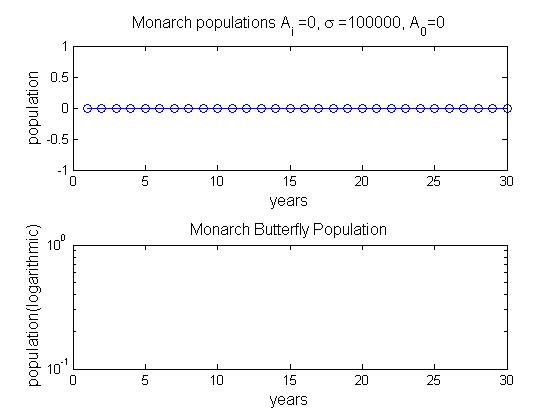}
\label{zeroandbig}
}
\caption{Large values of $\sigma$ lead to extinction.}
\end{figure}

To verify our model reflects reality, we simulated the extinction of the milkweed and in our model the monarch butterfly population dies off as expected.
If we simulate the population of milkweed without harvesting from herbicide, we see that the monarch butterfly population increases by a factor of 2000. 

\begin{figure}%
\centering
\subfigure[Annual monarch butterfly population behavior in the United States with $A_0 = 0$ and $\sigma = 100000$ per day]
{%
\includegraphics[width=7cm]{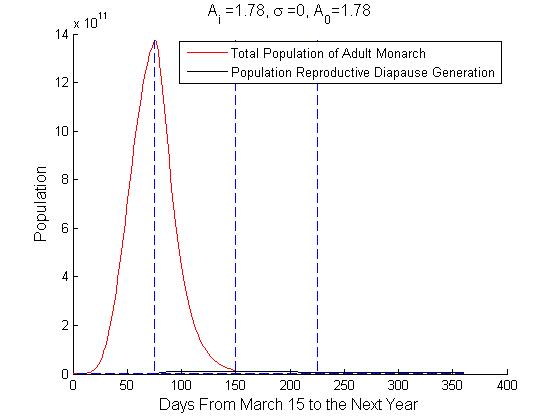}
\label{deadbutterflies}
}
\hspace{8pt}%
\subfigure[Prediction of the monarch butterfly population in the United States over a $30$ year period with $A_0 = 1.78$ and $\sigma = 0$]
{
\includegraphics[width=7cm]{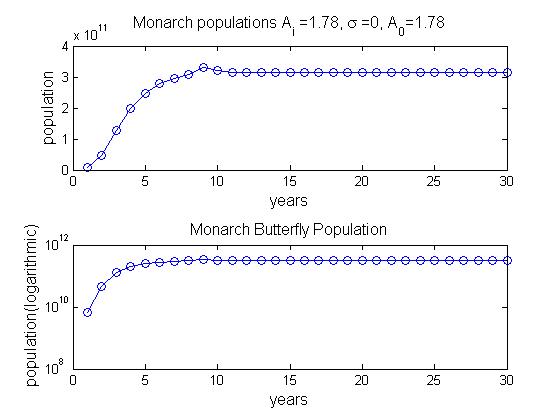}
\label{alivebutterflies2}
}%
\caption{Lack of herbicidal spraying leads to population stability.}
\end{figure}

\subfiglabelskip=0pt
\begin{figure}[htp]
\centering
\subfigure[Annual monarch buttery population behavior in the United States with $A_{0} = 0.4$, $A_{i} = 0.7$, and $\sigma = 1$ per day.]
{%
\includegraphics[width=7cm]{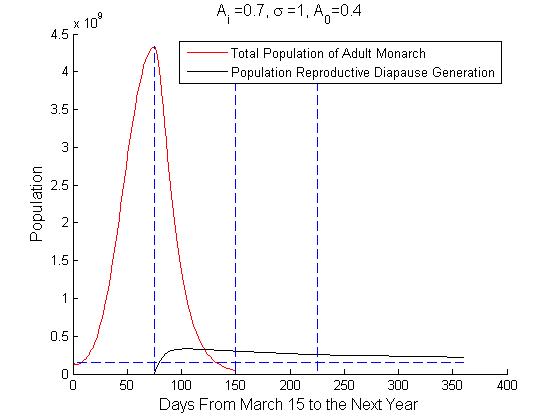}
\label{fig:canhigher}%
}
\hspace{8pt}%
\subfigure[Annual monarch buttery population behavior in the United States with $A_{0} = 0.7$, $A_{i} = 0.4$, $\sigma = 1$ per day.]{%
\includegraphics[width=7cm]{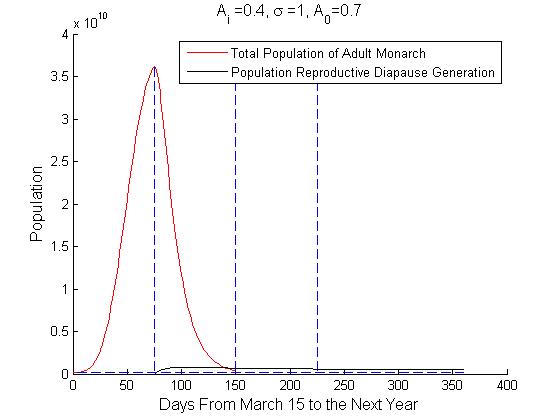}
\label{fig:ushigher1}%
}\\
\subfigure[Simulation of the monarch buttery population in the United States over a 30 year period with $A_{0} = 0.4$, $A_{i} = 0.7$, and $\sigma = 1$.]{%
\includegraphics[width=7cm]{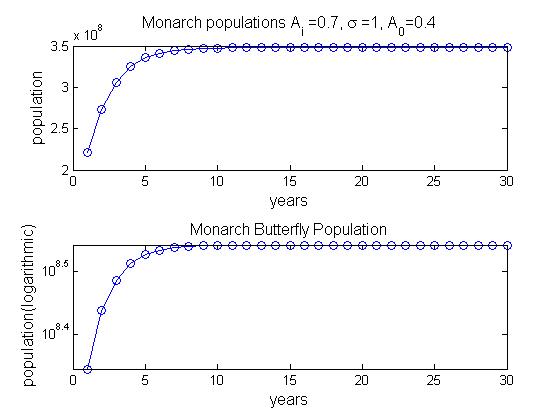}
\label{fig:canhigher2}%
}%
\hspace{8pt}%
\subfigure[Simulation of the monarch buttery population in the United States over a 30 year period with $A_{0} = 0.4$, $A_{i} = 0.7$, and $\sigma = 1$.]{%
\includegraphics[width=7cm]{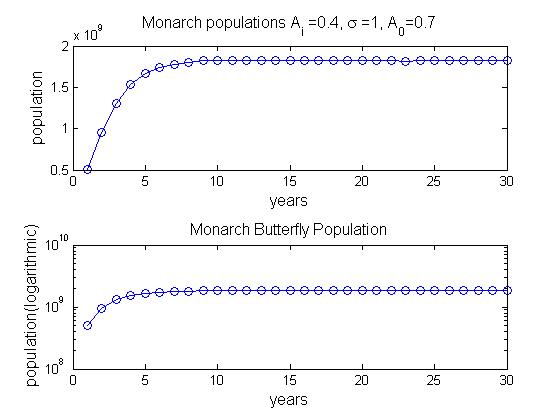}
\label{fig:ushigher2}%
}%
\caption{Milkweed population along the migratory route has significant effect on long term population size.}
\label{fig:diffinsig}%
\end{figure}

We vary the amount of milkweed at the various stages of the model, in the United States and Canada, in our simulations.  We see the monarch butterfly population is more sensitive to milkweed in the southern United States than other areas.  

\begin{figure}[htp]
\centering
\includegraphics[width=10cm]{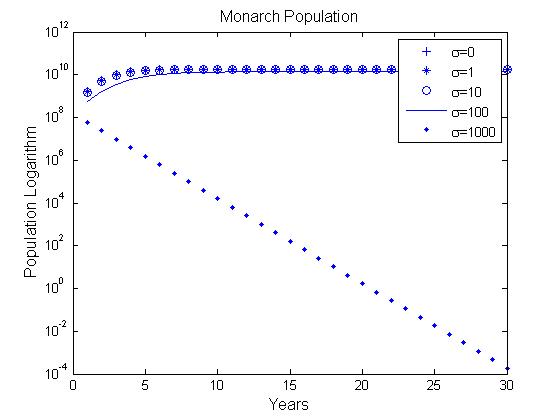}
\caption{Simulation of the monarch buttery population in the United States over a 30 year period with $A_{0} = 1$ and $A_{i} = 1$.
}
\label{fig:as1varys}
\end{figure}

For $A_0=1$, and $A_i=1$ there is not much difference when $\sigma = 0, 1, 10$ or $100$ but there is a difference when $\sigma = 1000$.
For instance, the first four values of $\sigma$, the value approaches $10^{10}$ on a logarithmic scale and when $\sigma=1000$ it approaches 0.
We note that when $\sigma=100$ it has a moderate effect on the population.

\begin{figure}[htp]
\centering
\includegraphics[width=10cm]{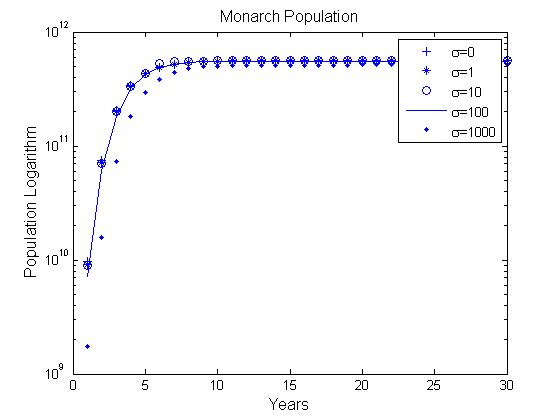}
\caption{Simulation of the monarch buttery population in the United States over a 30 year period with $A_{0} = 2$ and $A_{i} = 2$.}
\label{fig:as2varys}
\end{figure}

For $A_0=2$ and A$_i=2$  the graphs are similar, i.e. they have approximately the same upper horizontal asymptote, but when $\sigma$ varies the only  change is the shape. 

\begin{figure}[htp]
\centering
\includegraphics[width=10cm]{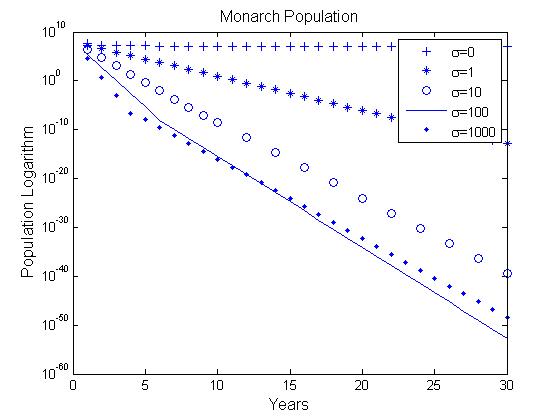}
\caption{Simulation of the monarch buttery population in the United States over a 30 year period with $A_{0} = 0.5$ and $A_{i} = 0.5$.}
\label{fig:aspoint2varys}
\end{figure}

For $A_0=0.2$ and $A_i=0.2$ we see when $\sigma=0$ the curve approaches a nonzero asymptote but when $\sigma = 1$, the curve approaches a zero asymptote, and when $\sigma = 10$, the population quickly dies off and so on with higher values of $\sigma$.
The curves are more sensitive to higher value of $\sigma$'s.

\begin{figure}[htp]
\centering
\includegraphics[width=10cm]{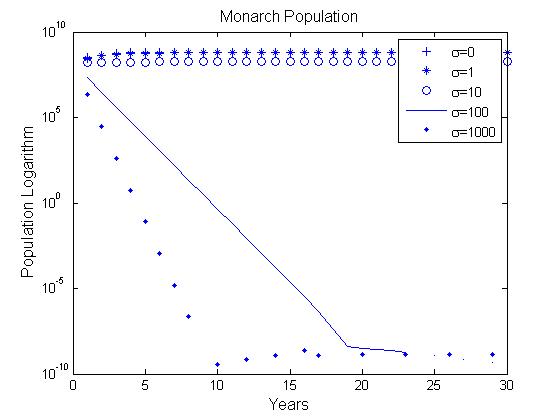}
\caption{Simulation of the monarch buttery population in the United States over a 30 year period with $A_{0} = 0.2$ and $A_{i} = 0.2$.
}
\label{fig:aspoint5varys}
\end{figure}

For $A_{0}=0.5$ and $A_{i}=0.5$, we observe that for $\sigma=0$, the curve has a large asymptotic value, but when $\sigma=1$ this asymptote is lower.  When $\sigma =10$ the curve decrease to about third of its initial value.
When $\sigma=100$ the value approaches zero.  In the simulation, we see that for $\sigma=1000$ it has oscillatory behavior.
This graph also shows another asymptote, which the monarch butterfly population tends to go to  $\sigma=10$.

\begin{figure}[htp]
\centering
\includegraphics[width=10cm]{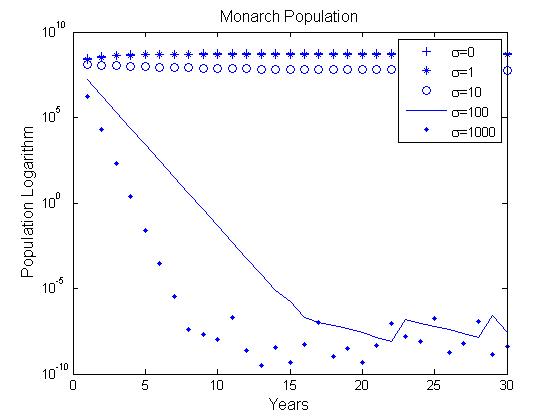}
\caption{Simulation of the monarch buttery population in the United States over a 30 year period with $A_{0} = 0.5$ and $A_{i} = 0.4$.
}
\label{fig:a0spoint5apoint4varys}
\end{figure}

For $A_0=0.5$ and $A_i=0.4$ and $\sigma=0$ the population converges to an asymptotic line.
For $\sigma = 1000$, the population approaches extinction.
Although this level of herbicide is excessive it is reasonable because with the amount of herbicide used, it will cause to population to go to zero.

\subsection{Estimation of Population Density}

Obtaining an accurate population count of all monarch butterflies is very difficult, if not impossible.
We attempt to find the population density at specific times, so that future experiments can either verify or refute our model.

We chose three coordinate points on the map of the United States to create three vertices of a triangle. This triangle covers the area where monarch butterfly activity was observed. The three points are: $A = (23 N, 100 W)$ at the central Mexico, $B = (48 N, 102 W)$ at the Northwestern U.S. and $C = (48 N, 71 W)$ at the Northeastern U.S. 

\begin{figure}[htp]
\centering
\includegraphics[width=7cm]{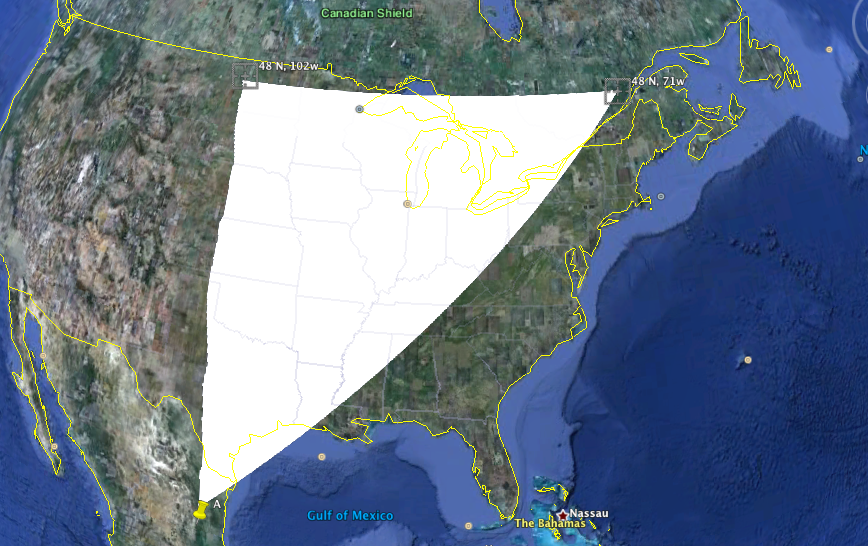}
\caption{An approximation of the area or reported monarch butterfly activity.}
\label{fig:map}
\end{figure}

In Figure \ref{fig:map}, the left side of the triangle lies along the eastern side of the Rocky Mountain range. 
We assume the area is a right triangle.
We consult Google Earth, an online geographical information system, to find the area of the triangle,approximately $3,111,352 \, \unit{km^2}$.

In order to calculate the density for each day, we divide the area into 75 horizontal segments with equal height, where each segment represents the distance traveled by the monarch butterflies in one day of the spring migration.
This is not a robust assumption, because the population is likely distributed in multiple strips. 

We index the numbers from 1 through 75, from South to North.
Based on the formula for the area of a right triangle, the formula for calculating the area of a particular strip is given as:

\begin{equation*}
Area(i) = \frac{Area_{total}}{75^2} \cdot (i^2-(i-1)^2) 
\end{equation*}

We presume the entire population of monarch butterflies in the $i^{th}$ strip at the $i^{th}$ day of the spring migration.
The $75^{th}$ strip represents the second stage.
The $(75-i)^{th}$ strip is at the $i^{th}$ day of the fall migration and finally the $1^{st}$ strip represents the forth stage.

\begin{figure}[htp]
\centering
\includegraphics[width=10cm]{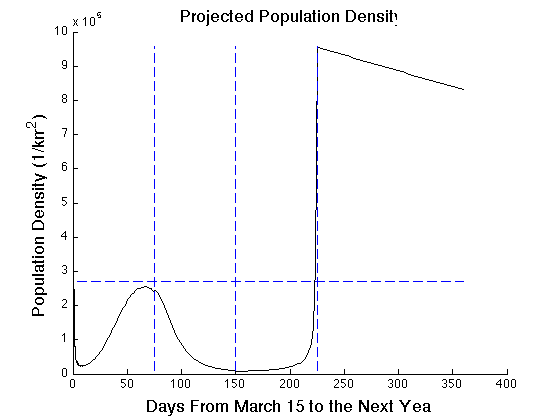}
\caption{population density of the monarch butterfly over four stages of migration}
\label{density}
\end{figure}

We calculate the population density at each specific time as shown in Figure \ref{density}.

\section{Discussion}

If we simulate the model over the course of a year, we observe unique dynamic behavior.
The monarch butterfly generations overlap one another and the generations grow with respect to time.
When the monarch butterflies arrive at Canada, we observe different behavior, there is a decline in the monarch and larvae population.
This is the behavior we observe in nature, because the monarch butterfly is stationary in Canada and they stay for an extended period of time (limited food resorces).
This means the milkweed population changes with respect to time.
When we repeat the simulation over many years the monarch butterfly population shows logistic growth.  

If we set the constants $A_0$ and $A(100)=A_i$ at a low enough value, then we see a change in behavior.
The graphs of the population invert with respect to time.
This suggests a critical point.
When we run simulations closer and closer to the critical point, we notice the range of values decrease.
When we look at this for one year, we see that higher values of $A$ exhibit logistic growth and lower values of $A$ exhibit logistic decay depending on the value of $\sigma$.

A sensitivity analysis of $\sigma$ shows that for small $\sigma$ the population is localized in one area. 
Increasing $\sigma$ by a factor of 10 will lead to extinction while smaller values of $\sigma$ show stabilization.  

We model the population when there is total extinction of the milkweed, to verify our model behaves as expected and the monarch butterfly population becomes extinct as well. 
We also look at the effect a value of $\sigma = 0$ has on the butterfly.
We see that the population increases by a factor of 2000, as time increases.     

We vary the values of the milkweed in the United States and in Canada through our simulations.
We see a larger impact on the overall population if there is more milkweed present in the United States than in Canada.
This difference is maybe due to a larger supply of milkweed during their migration and the larger portion of time spent in the United States.  

This value changes as we change the value of $\sigma$.
We conclude that herbicide has a large effect and a reduction of herbicidal spraying is needed to stabilize the monarch butterfly population.
In 2002, a severe winter storm in central Mexico caused the death of approximately 80\% of the monarch butterfly population, at the central Mexico overwintering site \cite{jnorth}.
The population rebounded the next year, though \cite{jnorth}.
There was likely enough milkweed for the monarch butterfly population to increase.
In later years it seems that the monarch butterfly population is smaller than the average value over the past several years \cite{jnorth}, indicating it may converge to a state.
Oscillating monarch population is due much to detrimental weather and declining forest population and we can only say that the herbicide has an effect on where this oscillation should be.

\section{Further Work}
\label{furtherwork}

For a more accurate picture of these simulations, more data collection sites located south of Cape May, NJ and west of Chincoteague, VA would be instructional.
These collection sites would be valuable because we know that during the fall migration, the monarch butterfly travels in a southwestern direction.  This means the data at all of these sites could be used to find an approximation of the population via method of numerical integration.
It would also be advantageous if Chincoteague revived their data collection simultaneous to the other site.
Not only should these three sites be operating at the same time, they should all be using the same method of collection.
Further research needs to be done on the population dynamics of milkweed to obtain a more accurate model of the butterfly population.
Another potential research topic related to dynamics is the effect of the Oyamel fir forest, in Michoac\'an, Mexico, on the monarch butterfly population.

\section{Acknowledgments}
\label{acknowledgment}

We thank Professor Carlos Castillo-Chávez, Professor Luis Melara, Professor Steve Wirkus, Professor Erika Camacho, Professor Baojun Song, Professor Bin Cheng, Susan Holechek, and Romarie Morales, Ben Morin, David Murillo for support, 
valuable discussions and encouragement. This research is supported in part by 
the National Science Foundation--Enhancing the Mathematical Sciences Workforce 
in the 21st Century (EMSW21), award \# 0838705; the Alfred P. Sloan Foundation-
-Sloan National Pipeline Program in the Mathematical and Statistical Sciences, 
award \# LTR 05/19/09; and the National Security Agency--Mathematical \& 
Theoretical Biology Institute---Research program for Undergraduates; award \# 
H98230-09-1-0104.

\nocite{MR1822695}
\bibliographystyle{acm}	
\bibliography{monarch}

\appendix
\section{Parameters for Stage 1}
\label{appstageone}

The monarch butterfly start out in central Mexico at a latitude of $20^{\circ}$ on the $75^{th}$ day of the year.
For calculation purposes we re-scale time by setting 75 to day 0.
Because the eggs are laid in the southern United States ($30^{\circ} - 35^{\circ}$) we can assume that 5\% of the adult monarchs will remain in this area.
Referring to the above figure the butterflies will reach $35^{\circ}$ approximately 25 days after departure.
The general solution to the first equation in our model is:
$$
 M_0=M_{0i} \cdot e^{-\mu_{0}t}
$$
Then to estimate our parameters we obtain:
$$
0.05 M_{0i}=M_{0i} e^{-25 \mu_0}
$$
by substitution
$$
\mu_0=-\log(0.05)/25   \approx 0.1198
$$
According to the Figure \ref{fig:monmig}, the butterflies will reach $30^{\circ}$ latitude 15 days after departure.
Then using the same approach as above we obtain $\mu_0=  0.1997$.
This gives us a range of values for $\mu_0$.

According to Urquhart, the average time for a larva to mature from egg to butterfly is approximately 28 days \cite{book4}.  This means there is a maturation rate  $\gamma=1/28$ per day.  
The parameter $\mu_1$ is calculated from the high mortality rates of the larvae.
From Dively et al 92 to 98 percent of the larvae do not make it to the adult stage.
Since it takes 28 days for the larva to mature 92 to 98 percent of the larvae population will die within 28 days.
Then using the general exponential solution we obtain values from 0.0902 to 0.1397 per time for our $\mu_1$.

A monarch female lays from 500 to 700 eggs over her lifespan \cite{book4} and her average life-span is approximately four weeks, or 28 days \cite{book4}.
This means that each each female monarch lays, on average, between 17 and 25 eggs each day, approximately 17.857 to 25 eggs.
This gives a reproductive rate of 8.929 to 12.5 per monarch butterfly.  Not all eggs are entirely fertile and the number of infertile eggs can be as high as 55\% \cite{book3}.
This gives a range of 4.91 and 12.5 larvae per monarch, which we use for our values of $\alpha_{1}$.

In \emph{stage 1}, the adult monarchs live anywhere from 2 to 6 weeks.
Thus to average $\mu_2$ we can assume the worst for these monarchs and assume that on average they live 2 weeks this means $\mu_2=1/14$ per day.

\section{Parameters for Stage 2}
\label{appstagetwo}

For the parameters in stage 2, we use some of the same parameters from \emph{stage 1}.
Because stage 2 is a stopping point, we vary the milkweed in the Equation \ref{stagetwoeqthree} which makes our $\alpha_2$ different from $\alpha_1$.
The parameter $\alpha_{2}$ is also different because the milkweed units have changed.
We assume a value of $2.6 \, \frac{\textrm{m}^2}{\textrm{kg} \cdot \textrm{day}}$ for $\alpha_{2}$, because the milkweed population will be decreasing not as a percentage but as in biomass thus $\alpha_{2} = 0 \, \frac{\textrm{m}^2}{\textrm{kg} \cdot \textrm{day}}$.

In stage 2 the butterflies in reproductive diapause are no longer mating and they are preparing for the upcoming journey by consuming more food.  
The  butterflies in the reproductive diapause state have not been traveling either.
This means the mortality rate $\mu_{3}$ will be a small value.
Thus we can assume that $\mu_3=0.005$ per day, since $1/\mu_{3} = 200$ days, which is approximately 6.7 months.

Due to insufficient data the carrying capacity of the common milkweed could not be found, but there did exist sufficient information on the butterfly milkweed \cite{}.
According to the grower's guide the optimal density of milkweeds was grown in the field.
This included a density of 43,560 plants per acre.
The grower's guide also included the dry weight herb of 104.7 $\frac{\textrm{g}}{\textrm{plant}}$ and a dry weight root of 61.9 $\frac{\textrm{g}}{\textrm{plant}}$.
Then the weight of the entire plant is:
$$
104.7 + 61.9=166.6 \, \frac{\textrm{g}}{\textrm{plant}}
$$
To find carrying capacity we multiply:
$$
166.6 \, \frac{\textrm{g}}{\textrm{plant}} \cdot 43560 \, \frac{\textrm{plants}}{\textrm{A}} \cdot 1 \frac{\textrm{A}}{4050 \textrm{m}^2} \cdot 1 \, \frac{\textrm{kg}}{1000\textrm{g}} = 1.79188 \, \frac{\textrm{kg}/}{\textrm{m}^2}.
$$
The growth rate, $a$, of the milkweed was also obtained from the data.
It appears that the total mass of a year one plant using similar calculations as before is 10.6 g/plant and that of a year two plant is 132.41g/plant.
To calculate the growth rate, we use exponential growth once more.  Let the initial condition be 10.6, then we can substitute in the new values in:
$$
A=A_1 e^{at}
$$
$$
132.41 = 10.6 \cdot e^{365a} \textrm{ by substitution}
$$
$$
a = 0.007 \textrm{ per day.}
$$
For $\beta$ we know that the value of the larvae population will be on the order of $10^8$.  Because we know that there should not be a high decline rate of the milkweed we can assume $\beta=5 \cdot 10^-9$.

\section{Parameters for Stage 3 and 4}
\label{appstage3and4}

Upon the return trip to central Mexico an estimation of the parameter came from the tagged data from the Monarch Watch website data base \cite{jnorth}.
The data was filtered out through the process of having non-dated taggings removed and considering only monarchs that were tagged after August 8th for any year.
This is important because we need to only consider those in migration.
Then out of the ones that we are considering, the amount that made it to Mexico.
It was found that 84 percent of the monarch’s made it.
Then we can say 16 percent of the migrants die on their journey south.
From August 15 to to November 1 we have a time span of about 75 days.
This gives a value of $\mu_{fin}=0.002325$ per day.

Then from the Journey North website it is stated that during their stay at the overwintering site approximately 15 percent of the population dies off due to predation.
Their stay at the overwintering sight from November 1 to March 15 is about 135 days.
Then by the same method a before we obtain as parameter estimate of $\mu_w=0.001204$.  

\section{MATLAB CODE: testa.m}
\begin{verbatim}
function [x,pop] = testa(P,N,M_0,A_i,P_2,P_3,P_4,flag)
% The function outputs the ending population of normal adult of stage 1 and
% of the reproductively diapaused adults of stage 2,3 and 4 as a row vector x.
% The input includes the arguments P, P_2, p_3, P_4 for the parameters of 
% differential equations of stages 1, 2, 3 and 4 respectively. flag determines 
% weather the function outputs graph or not, its default value is 1.

% example possible input:  testa([0.1198,2.6,.6,1/28,0.10147,1/14],4,150000000, ...
% 0.6,[2.6,0.0025,0.007,1.78788552,1,.000000007],.002298,0.001024,1);

%STAGE 1%%%%%%%%%%%%%%%%%%%%%%%%%%%%%%%%%%%%%%%%%%%%%%%%%%%%%%%%%%%%%%%%%%%
tspan=[0,75]; %time span of stage 1

y0(1)= M_0;
for i = 2:1:2*N+1 %set initial condition for nth generation of larvae and adult
    y0(i)=0;
end

[t,y]=ode45(@monarch,tspan,y0,[],P); %solve system of equations of stage 1

    function yprime=monarch(t,y,p) %define equations of stage 1
        mu_0=p(1);
        alpha_1=p(2);
        A_0=p(3);
        gamma=p(4);
        mu_1=p(5);
        mu_2=p(6);
        
        yprime(1)=-mu_0*y(1); %define equations cyclicly
        
        for j=1:N
            yprime(2*j)=alpha_1*y(2*j-1)*A_0-(mu_1+gamma)*y(2*j);
            yprime(2*j+1)=gamma*y(2*j)-mu_2*y(2*j+1);
        end
        
        yprime=transpose(yprime); % since yprime is defined as row vector, 
                                  % it has to be transposed to be used as column vector
    end

%KNITTING STAGE 1 & STAGE 2 %%%%%%%%%%%%%%%%%%%%%%%%%%%%%%%%%%%%%%%%%%%%%%%

for i = 1:N %creating a matrix that stores i columns of generations' larvae population
    l1Total(i) =  y(end,2*i);
end

for i = 1:N %creating a matrix that stores i columns of generations' adult population
    m1Total(i) = y(end,2*i+1);
end

%STAGE 2%%%%%%%%%%%%%%%%%%%%%%%%%%%%%%%%%%%%%%%%%%%%%%%%%%%%%%%%%%%%%%%%%%%
z0 = [sum(l1Total);sum(m1Total);0;0;A_i]; %sum all the entries of each column of the matrice, making it a row vector
tspan_2 = [0,75]; %time span of stage 2
[t_2,z]=ode15s(@monarch_2,tspan_2,z0,[],[P,P_2]);

    function zprime = monarch_2(t_2,z,p)
        
        mu_0=p(1);
        alpha_1=p(2);
        A_0=p(3);
        gamma=p(4);
        mu_1=p(5);
        mu_2=p(6);
        
        alpha_2=p(7);
        mu_3=p(8);
        a=p(9);
        K=p(10);
        sigma=p(11);
        beta=p(12);
        
        zprime(1)= -(gamma+mu_1)*z(1); %L_in
        zprime(2)= -mu_2*z(2)+gamma*z(1); %M_in
        zprime(3)=alpha_2*z(2)*z(5)-(gamma+mu_1)*z(3); %L_reproductive Diapause
        zprime(4)=gamma*z(3)-mu_3*z(4); %M_Reproductive Diapause
        zprime(5)=a*z(5)*(1-z(5)/K)-z(5)*(sigma+beta*z(3)); %A
        
        zprime = transpose(zprime);
    end

%KNITTING STAGE 2 & STAGE 3 %%%%%%%%%%%%%%%%%%%%%%%%%%%%%%%%%%%%%%%%%%%%%%%

m2Total = z(end,4); %take the final value of adult population

%STAGE 3%%%%%%%%%%%%%%%%%%%%%%%%%%%%%%%%%%%%%%%%%%%%%%%%%%%%%%%%%%%%%%%%%%%
v0 = m2Total;
tspan_3=[0,75];
[t_3,v]=ode45(@monarch_3,tspan_3,v0,[],P_3);

    function vprime = monarch_3(t_3,v,p)
        mu_fin = p(1);
        vprime(1) = -mu_fin*v(1);
    end
% plot(t_3,v);
% v(end)

%KNITTING STAGE 3 & STAGE 4 %%%%%%%%%%%%%%%%%%%%%%%%%%%%%%%%%%%%%%%%%%%%%%%
m3Total = v(end);

%STAGE 4%%%%%%%%%%%%%%%%%%%%%%%%%%%%%%%%%%%%%%%%%%%%%%%%%%%%%%%%%%%%%%%%%%%
w0 = m3Total;
tspan_4 =[0,135];
[t_4,w]=ode45(@monarch_4,tspan_4,w0,[],P_4);

    function wprime = monarch_4(t_4,w,p)
        mu_w = p(1);
        wprime(1) = -mu_w*w(1);
    end
% plot(t_4,w);

%PLOTTING %%%%%%%%%%%%%%%%%%%%%%%%%%%%%%%%%%%%%%%%%%%%%%%%%%%%%%%%%%%%%%%%%

ta = tspan(2); tb = tspan_2(2); tc = tspan_3(2); td = tspan_4(2);
%store the time period (day) of each stage

if nargin < 8 %setting default value of flag
    flag = 1;
end

for i=1:N+1
    b_total(:,i)=y(:,2*i-1);
%creating a matrix containing N generations of adult population
end

for i=1:N
    l_total(:,i)=y(:,2*i);
%creating a matrix with N generations of larvae population
end

if (flag) %making this function output graphs only if flag != 0
    
    %first graph with all the curves of 4 stages in normal scale
    
    figure;
    
    hold on;
    
    %Performing for-loops in the if-statement again, otherwise
    %this if-statement comes up with errors. And it will comes up with
    %b_total undefined error if thess for-loops are only instead the
    %if-statement. Dont know why, I guess that the scopes of variables may
    %span differently than that in Java.
    
    for i=1:N+1
        curve1_1=plot(t,y(:,2*i-1),'b'); 
%plotting each generation of larvae population
    end
    
    for i=1:N
        curve1_2=plot(t,y(:,2*i),'g'); 
%plotting each generation of adult population
    end
    
    for i=1:N+1
        b_total(:,i)=y(:,2*i-1); 
%creating a matrix containing N generations of adult population
    end
    
    for i=1:N
        l_total(:,i)=y(:,2*i); 
%creating a matrix with N generations of larvae population
    end
    curve1_3=plot(t,sum(l_total'),'black'); 
% transposing the matrix, and summing each entry in each column, making 
% it a row vector and plotting it
    plot(t_2+ta,z(:,1),'black'); 
%plotting population of larvae_in at stage 2
    
    curve1_4=plot(t,sum(b_total'),'r'); 
%plotting total population of monarch_in at stage 1
    plot(t_2+ta,z(:,2),'r');            
%plotting population of monarch_in at stage 2
    
    curve1_5=plot(t_2+ta,z(:,3),'magenta'); 
%plotting stage 2 super larvae population
    curve1_6=plot(t_2+ta,z(:,4),'cyan'); 
%plotting stage 2 super monarch population
    
    plot(t_3+ta+tb,v,'cyan'); 
%plotting stage 3 super monarch population
    plot(t_4+ta+tb+tc,w,'cyan'); 
%plotting stage 4 super monarch population
    
    xtotal = ta+tb+tc+td;
    plot([0,xtotal],[M_0,M_0],'--'); 
%plotting a horizontal line with vertical value M_0
    
    plot([ta ta],[0 max(sum(l_total'))],'--'); 
%plotting a vertical bar that separates stage 1&2
    plot([ta+tb ta+tb],[0 max(sum(l_total'))],'--'); 
%plotting a vertical bar that seperates stage 2&3
    plot([ta+tb+tc ta+tb+tc],[0 max(sum(l_total'))],'--'); 
%plotting a vertical bar that seperates stage 3&4
    
    legend([curve1_1,curve1_2,curve1_3,curve1_4,curve1_5,curve1_6],'Population of ...
        adult of each generation','Population of larvae of each generation','Total ...
        population of adult','Total population of larvae','Population of next ....
        generation larvae', 'Population of reproductive diapause generation');
    
    title_1=num2str(A_i);
    title_2=num2str(P_2(5));
    title_3=num2str(P(3));
    d1=title(strcat('A_i = ',title_1,', sigma =',title_2,', A_0= ',title_3));
    d2=xlabel('Years');
    d3=ylabel('Population');
    set(d1,'FontSize',12);
    set(d2,'FontSize',12);
    set(d3,'FontSize',12);
    
    figure;
    %graph only the population of adult butterfly over 4 stages.
    
    hold on;
    
    curve2_1=plot(t,sum(b_total'),'r');
    plot(t_2+ta,z(:,2),'r');
    curve2_2=plot(t_2+ta,z(:,4),'black');
    plot(t_3+ta+tb,v,'black');
    plot(t_4+ta+tb+tc,w,'black');
    plot([ta ta],[0 max(sum(b_total'))],'--');
    plot([ta+tb ta+tb],[0 max(sum(b_total'))],'--');
    plot([ta+tb+tc ta+tb+tc],[0 max(sum(b_total'))],'--');
    plot([0,ta+tb+tc+td],[M_0,M_0],'--');
    
    legend([curve2_1,curve2_2],'Total Population of Adult Monarch',...
        'Population Reproductive Diapause Generation');
    
    a1=title(strcat('A_i = ',title_1,', sigma =',title_2,', A_0= ',title_3));
    a2=xlabel('Days From March 15 to the Next Year');
    a3=ylabel('Population');
    set(a1,'FontSize',12);
    set(a2,'FontSize',12);
    set(a3,'FontSize',12);
        
end
%SETTING OUTPUT OF THE FUNCTION%%%%%%%%%%%%%%%%%%%%%%%%%%%%%%%%%%%%%%%%%%%%%%%%
x(1) = y(end,2*N+1); %ending value of normal adult of stage 1
x(2) = z(end,4);     %ending value of super adult of stage 2
x(3) = v(end);       %ending value of super adult of stage 3
x(4) = w(end);       %ending value of super adult of stage 4

%LATER MODIFICATION%%%%%%%%%%%%%%%%%%%%%%%%%%%%%%%%%%%%%%%%%%%%%%%%%%%%%%%%

temp_1 = vertcat(t,ta+t_2,t_3+ta+tb,t_4+ta+tb+tc); 
%concatenating time output
temp_2 = vertcat((sum(b_total'))',(sum(vertcat(z(:,4)',z(:,2)')))',v,w); 
%concatenating population output
pop = horzcat(temp_1,temp_2); 
%outputs two columns, with each contains time vector and adultpopulation vector

end
\end{verbatim}
\section{MATLAB CODE: visual.m}
\begin{verbatim}
a=xlsread('milkweedData.xlsx');
%reads the excel file
b=xlsread('FirstMonarch2010.xlsx');
c=xlsread('OtherMonarch2010.xlsx');
d=xlsread('stuffFor2011.xlsx');

%%%%%%%%%plots the figures in the paper%%%%%%%%%

%plot((a(:,4)-1)*30+a(:,3),a(:,1),'r.',(d(:,4)-1)*30+d(:,3),d(:,1),'b.');
a1=xlabel('Day of the Year 2011');
a2=ylabel('Latitude');
legend('First Milkweed Leaves','First Adult Monarch Sighting');
set(a1,'FontSize',12);
set(a2,'FontSize',12);

%%%%%%%%%plots the other figure in the paper%%%%%%%%%

%plot((b(:,4)-1)*30+b(:,3),b(:,1),'b.',(c(:,4)-1)*30+c(:,3),c(:,1),'r.');
%a1=xlabel('Day of the Year 2010');
%a2=ylabel('Latitude');
%legend('First Monarchs','Other Monarchs');
%set(a1,'FontSize',12);
%set(a2,'FontSize',12);
\end{verbatim}
\section{MATLAB CODE: density.m}
\begin{verbatim}
function y=density(P,flag)
%This function outputs a graph of population density against time (day) in
%period of a year. P is a row vector contains all the parameters used in
%system of equations in the 4 stages migration. 

A_t = 3111352;
%the area of a right triangle that covers the area where activities of 
%monarch were observed.area is in unit square km.

%creating an array of two columns with the first column storing the date
%and second storing the area of strip of the triangle at the respective
%date. 

for i = 1:360; 
    A(i,1)= i;
end

%day 1, the area is the one that at central mexico.
A(1,2)=A_t/75^2;

%day 2 - 75, the areas are calculated by the folmula for the area of right
%triangle.
for i=2:75
    A(i,2)= A_t*((i/75)^2-((i-1)/75)^2);
end

%stage 2, the range of monarch activity stays at the same area
for i = 76:150
    A(i,2)=A_t*(1-(74/75)^2);
end

%stage 3, area decreases
for i = 151:225
    A(i,2) = A((226-i),2);
end

%stage 4, staying at central mexico
for i = 226:360
    A(i,2) = A(1,2);
end

P_1 = P(1:6);
N = P(7);
M_0 = P(8);
A_i = P(9);
P_2 = P(10:15);
P_3 = P(16);
P_4 = P(17);

[a,b]=testa(P_1,N,M_0,A_i,P_2,P_3,P_4,0);

%figure out the number of data points of a column of b;
N = numel(b(:,1));

for i=1:N
    %round the time to the nearest integer
    temp = ceil(b(i,1));
    if (temp == 0)
        area(i)=A(1,2);
    else
        area(i)=A(temp,2);
    end
    
    %calculating density
    D(i) = b(i,2)/area(i);
end

if (flag==1)
    figure;
    hold on;
    plot(b(:,1),D,'black');
    plot([0,360],[D(1),D(1)],'--');
    a1=xlabel('Days From March 15 to the Next Year');
    a2=ylabel('Population Density (1/km^2)');
    a3=title('Projected Population Density');
    plot([75 75],[0 max(D)],'--');
    plot([150 150],[0 max(D)],'--');
    plot([225 225],[0 max(D)],'--');
    set(a1,'FontSize',12);
    set(a2,'FontSize',12);
    set(a3,'FontSize',12);
end
end
\end{verbatim}

\section{MATLAB CODE: endingV.m}
\begin{verbatim}
function [y]=endingV (Y,P,flag)
%Y time period in unit of year for the function to simulate
%p parameters for function testa.m
%flag optional variable, flag = 0 means no output graph

%example possible code: endingV(30,[[0.1198,2.6,.6,1/28,0.10147,1/14],4,150000000,0.6,[2.6,0.0025,0.007,1.78788552,1,.000000007],.002298,0.001024],1);
 
P_1 = P(1:6);
N = P(7);
M_0(1) = P(8);
A_i = P(9);
P_2 = P(10:15);
P_3 = P(16);
P_4 = P(17);

for i = 1:Y
[x(i,:),pop] = testa(P_1,N,M_0(i),A_i,P_2,P_3,P_4,0); %storing the results of each run
% of testa.m function into a row, and creating a matrix that piles these rows together
M_0(i+1) = x(i, 4); %Storing the final adult population to be used as initial M_0 for
%the next run.

end

if (nargin<3)
    flag = 1;
end

    title_1=num2str(A_i);
    title_2=num2str(P_2(5));
    title_3=num2str(P(3));

if (flag)
    figure;
    
    %Plotting On a logarithmic Scale
    
    subplot(2,1,1);
    plot(x(:,4),'-o'); %plotting final adult population in normal scale
    a3=title(strcat('Monarch populations A_i = ',title_1,', sigma = ',title_2,', ...
        A_0= ',title_3));
    a1=xlabel('years');
    a2=ylabel('population'); 
   
    set(a1,'FontSize',13);
    set(a2,'FontSize',13);
    set(a3,'FontSize',13);
       
    subplot(2,1,2);
    semilogy(x(:,4),'-o'); %plotting the same thing in log scale
    b1=title('Monarch Butterfly Population');
    b2=xlabel('years');
    b3=ylabel('population(logarithmic)');
    set(b1,'FontSize',12);
    set(b2,'FontSize',12);
    set(b3,'FontSize',12);
    
end

y=x(:,4);
end
\end{verbatim}
\section{MATLAB CODE: finaltemp.m}
\begin{verbatim}
a=xlsread('FirstMonarch2010withtemp.xlsx');%reads the excel file
plot((a(:,4)-1)*30+a(:,3),a(:,7),'b.');%temp vs days
a1=xlabel('Days of the Year');%Labels
a2=ylabel('Temperature in Celsius');
a3=title('Temperatures at First Sightings of Monarchs in 2010')
set(a1,'FontSize',12);
set(a2,'FontSize',12);
set(a3,'FontSize',12);
\end{verbatim}
\section{MATLAB CODE: refine.m}
\begin{verbatim}
function l=refine(a_0,a_i,sigma_i)
%This function plots the long term behavior of the monarch butterfly population but
%can plot more than one according to the parameter being considered and can
%label the parameter accordingly.
%One example is refine([1,1,1],[2,2,2],[0,0,0]
clear figure
[q,r]=size(a_0);
Y=30;%Default value for the number of years
flag=1;

for j=1:r
    %Default set for our parameters
    P=[[0.1198,2.6,a_0(j),1/28,0.10147,1/14],4,150000000,a_i(j),[2.6,0.0025,0.007, ...
        1.78788552,sigma_i(j),.000000007],.002298,0.001024];
    m(j,:)=endingV(Y,P,0);
    %semilogy(y(:,j));
end

hold all
for h=1:r
    semilogy(m(h,:),'.');%plots it on a logarithmic scale
    s{h}=strcat('sigma=',num2str(sigma_i(h)));%Display the sigma
    %s{h}=strcat('A_0=',num2str(a_0(h)));%Display milkweed in US
    %s{h}=strcat('A_i=',num2str(a_i(h)));%Display initial milkweed in
    %Canada
    legend(s);
end
c1=title('Long Term Trends on the Monarch Butterfly Population')
c2=xlabel('Years');
c3=ylabel('Population');
set(c1,'FontSize',12);
set(c2,'FontSize',12);
set(c3,'FontSize',12);
\end{verbatim}

\end{document}